\documentclass[a4paper,fleqn,usenatbib]{mnras}

\usepackage[T1]{fontenc}
\usepackage{ae,aecompl}

\usepackage{graphicx}	
\usepackage{amsmath}	
\usepackage{amssymb}	

\usepackage{ulem}
\usepackage[usenames]{color}
\definecolor{rred}{rgb}{0.8,0,0}

\title[LAT observations of the 2015 outburst of V404 Cyg]{High-energy gamma-ray observations of the accreting black hole V404 Cygni during its June 2015 outburst}

\author[Loh et al. ]
{A. Loh$^{1}$\thanks{E-mail: alan.loh@cea.fr},
S. Corbel$^{1, 2}$,
G. Dubus$^{3, 4}$,
J. Rodriguez$^{1}$,
I. Grenier$^{1}$,
T. Hovatta$^{5, 6}$,
\newauthor
T. Pearson$^{7}$,
A. Readhead$^{7}$,
R. Fender$^{8}$
and K. Mooley$^{8}$
\\
$^{1}$Laboratoire AIM (CEA/IRFU - CNRS/INSU - Univ. Paris Diderot), CEA DSM/IRFU/SAp, F-91191 Gif-sur-Yvette, France\\
$^{2}$Station de Radioastronomie de Nan\c{c}ay, Observatoire de Paris, PSL Research University, CNRS, Univ. Orl\'{e}ans, 18330 Nan\c{c}ay, France \\
$^{3}$Univ. Grenoble Alpes, IPAG, F-38000 Grenoble, France\\
$^{4}$CNRS, IPAG, F-38000 Grenoble, France\\
$^{5}$Aalto University Mets\"ahovi Radio Observatory, Mets\"ahovintie 114, 02540 Kylm\"al\"a, Finland\\
$^{6}$Aalto University Department of Radio Science and Engineering, P.O. BOX 13000, FI-00076 AALTO, Finland\\
$^{7}$California Institute of Technology, 1200 E. California Blvd, MC 249-17, Pasadena, CA 91125, USA\\
$^{8}$Astrophysics, Department of Physics, University of Oxford, Keble Road, Oxford OX1 3RH, UK
}

\date{Accepted 2016 July 14. Received 2016 July 13; in original form 2016 April 22}

\pubyear{2016}

\begin{document}
\label{firstpage}
\pagerange{\pageref{firstpage}--\pageref{lastpage}}
\maketitle

\begin{abstract}
We report on \textit{Fermi}/Large Area Telescope observations of the accreting black hole low-mass X-ray binary V404\,Cygni during its outburst in June--July 2015.  
Detailed analyses reveal a possible excess of $\gamma$-ray emission on 26 June 2015, with a very soft spectrum above $100$\,MeV, at a position consistent with the direction of V404 Cyg (within the $95\%$ confidence region and a chance probability of $4 \times 10^{-4}$).
This emission cannot be associated with any previously-known \textit{Fermi} source. 
Its temporal coincidence with the brightest radio and hard X-ray flare in the lightcurve of V404\,Cyg, at the end of the main active phase of its outburst, strengthens the association with V404\,Cyg. 
If the $\gamma$-ray emission is associated with V404 Cyg, the simultaneous detection of $511\,$keV annihilation emission by INTEGRAL requires that the high-energy $\gamma$ rays originate away from the corona, possibly in a Blandford-Znajek jet. 
The data give support to models involving a magnetically-arrested disk where a bright $\gamma$-ray jet can re-form after the occurrence of a major transient ejection seen in the radio. 

\end{abstract}




\section{Introduction}

\begin{figure}
\includegraphics[scale=1]{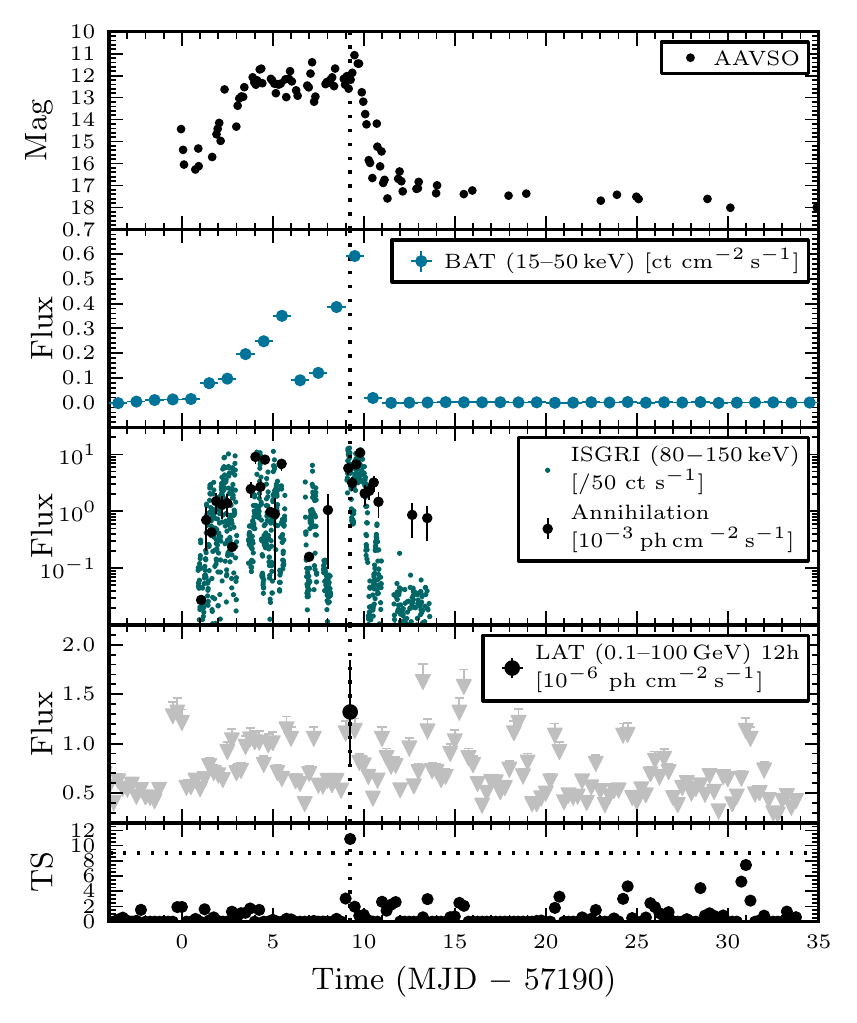}
\vspace{-10pt}
\caption{$\gamma$-ray flux and $95\%$ upper limits ${>}100\,$MeV and corresponding TS at the position of V404$\,$Cyg in 12-h bins, shifted by $6$ hours in time. The upper panels display the optical (extracted from the AAVSO database), hard X-ray \textit{Swift}/BAT \citep{2013ApJS..209...14K}, INTEGRAL/ISGRI count rate in the $80$--$150$\,keV band and the photon flux in the annihilation line \citep{2016Natur.531..341S} LCs.
\label{fig:lc_12h}}
\end{figure}	

The microquasars consist of an accreting black hole or neutron star in a binary system with transient or persistent relativistic jets \citep{1999ARA&A..37..409M}. 
They display a wide range of behaviour at all wavelengths \citep[e.g][]{2006csxs.book..381F}, but they have rarely been detected at high-energy $\gamma$ rays, despite the high-energy particles produced in their jets \citep{2002Sci...298..196C}. 
Relativistic particles in the jet could emit $\gamma$ rays, either by Compton up-scattering the low-energy photons from the accretion disc/stellar field \citep[e.g.][]{2002A&A...388L..25G, 2006A&A...447..263B} or by pion decays from inelastic collisions between jet particles and stellar wind protons \citep{2003A&A...410L...1R}. 
Gamma-ray photons could also be produced at the shock regions where the jets encounter the interstellar medium \citep{2011A&A...528A..89B} or within the jet itself \citep{1999MNRAS.302..253A}.
Despite their recurrent outbursts, only the microquasar Cyg X$-$3 \citep[and perhaps Cyg X$-$1,][]{2013ApJ...775...98B, 2013MNRAS.434.2380M} behaves like a clearly identified transient high-energy emitter \citep{2009Sci...326.1512F, 2009Natur.462..620T}.
The latter's $\gamma$-ray flares are strongly correlated to the radio emission originating from the relativistic jets \citep{2012MNRAS.421.2947C}. 
With a donor star of mass ${>}10\,\mbox{M}_{\sun}$, they are consistent with being  high-mass X-ray binaries.

The low-mass X-ray binary (LMXB) V404$\,$Cygni (also known as GS$\,$2023$+$338) underwent an exceptional outburst phase in June 2015. V404$\,$Cyg is a nearby  system \citep[$2.39 \pm 0.14\,$kpc, ][]{2009ApJ...706L.230M} harbouring a ${\sim}9\,\mbox{M}_{\sun}$ black hole and a ${\sim}1\,\mbox{M}_{\sun}$ companion star \citep{2010ApJ...716.1105K, 1992ApJ...401L..97W} with a  $6.5$-day orbital period \citep{1992Natur.355..614C}.
After a $26$ year long quiescent period \citep[{e.g.}][]{2016ApJ...821..103R}, renewed activity was detected with the \textit{Swift}/BAT \citep{2015GCN..17929...1B} and \textit{Fermi}/GBM \citep[][up to $300\,$keV]{2015GCN..17932...1Y, 2016arXiv160100911J} on June, 15 (MJD 57188). This triggered a worldwide multi-wavelength monitoring campaign from radio \citep[][]{2015ATel.7658....1M} to hard X-rays \citep[][]{2015A&A...581L...9R} until V404$\,$Cyg faded towards its quiescent accretion level in August 2015 \citep{2015ATel.7959....1S}.
The source has since undergone a fainter re-brightening, in December 2015 \citep[{e.g.}][]{2015ATel.8455....1B}.

We took advantage of the intense activity and monitoring of V404$\,$Cyg to probe its high-energy emission with the Large Area Telescope \citep[LAT,][]{2009ApJ...697.1071A} on board \textit{Fermi}.
The $\gamma$-ray analysis leading to the marginal detection of a flare is described in \S\ref{sec:analysis}. Its origin is discussed in \S\ref{sec:discussion}.

\section{Observations and data analysis} \label{sec:analysis}

We have analysed the Pass~8 \textit{Fermi}/LAT data covering the period 4 August 2008 to 17 July 2015. 
We used the \textit{Fermi} Science Tools (version v10r1p1) with the Instrument Response Functions set \texttt{P8R2\_SOURCE\_V6}. 
A $15\degr$ acceptance cone centered on the infrared position of V404\,Cyg has been considered. LAT photons labelled as \texttt{SOURCE} (\texttt{evclass=128}) were selected in the energy range from $100\,$MeV to $100\,$GeV.
Events were also filtered based on the quality of the PSF, choosing the 3 best partitions (PSF 1 to 3: \texttt{evtype=56}) and standard filters on the data quality were applied.
To minimise the contamination by Earth limb photons, $\gamma$-ray events with reconstructed directions pointing above a $90\degr$ zenith angle have been excluded.

In order to first constrain the emission of the whole region (nearby point-like sources and diffuse sky components),  we performed a \texttt{binned} maximum-likelihood spectral analysis using the \texttt{NewMinuit} optimization algorithm implemented in \texttt{gtlike}. 
In the modelling of the region of interest (RoI), we have included the standard templates for the Galactic and isotropic backgrounds 
(\texttt{gll\_iem\_v06.fits} and \texttt{iso\_P8R2\_SOURCE\_V6\_v06.txt}\footnote{\href{http://fermi.gsfc.nasa.gov/ssc/data/analysis/}{http://fermi.gsfc.nasa.gov/ssc/data/analysis/}} ) 
and we used the spectral models listed in the 4-year \textit{Fermi} catalogue \citep[][hereafter 3FGL]{2015ApJS..218...23A} for all the sources within a $25\degr$ radius. 
The normalisation parameters were left free to vary for the diffuse components and for the sources either within $5\degr$ from the RoI center or labelled as variable ({i.e.}\ a variability index greater than $72.44$, 3FGL). 
The results of the \texttt{binned} likelihood analysis using $7$~years of LAT data, without including V404\,Cyg which is not detected over this time scale, are fully consistent with those of the 3FGL catalogue. 
For instance, the flux normalisation for the blazar B2023$+$336 \citep[3FGL J2025.2$+$3340,][]{2012ApJ...746..159K}, which is only ${\sim}0\fdg32$ away from V404\,Cyg, differs by only ${\sim} 3\%$ from the catalogue value. In addition, the residual maps do not show any noticeable structures, especially around the Galactic plane. 

	\subsection{Gamma-ray variability} \label{sec:analysis_lc}
	
\begin{figure}
\includegraphics[scale=1]{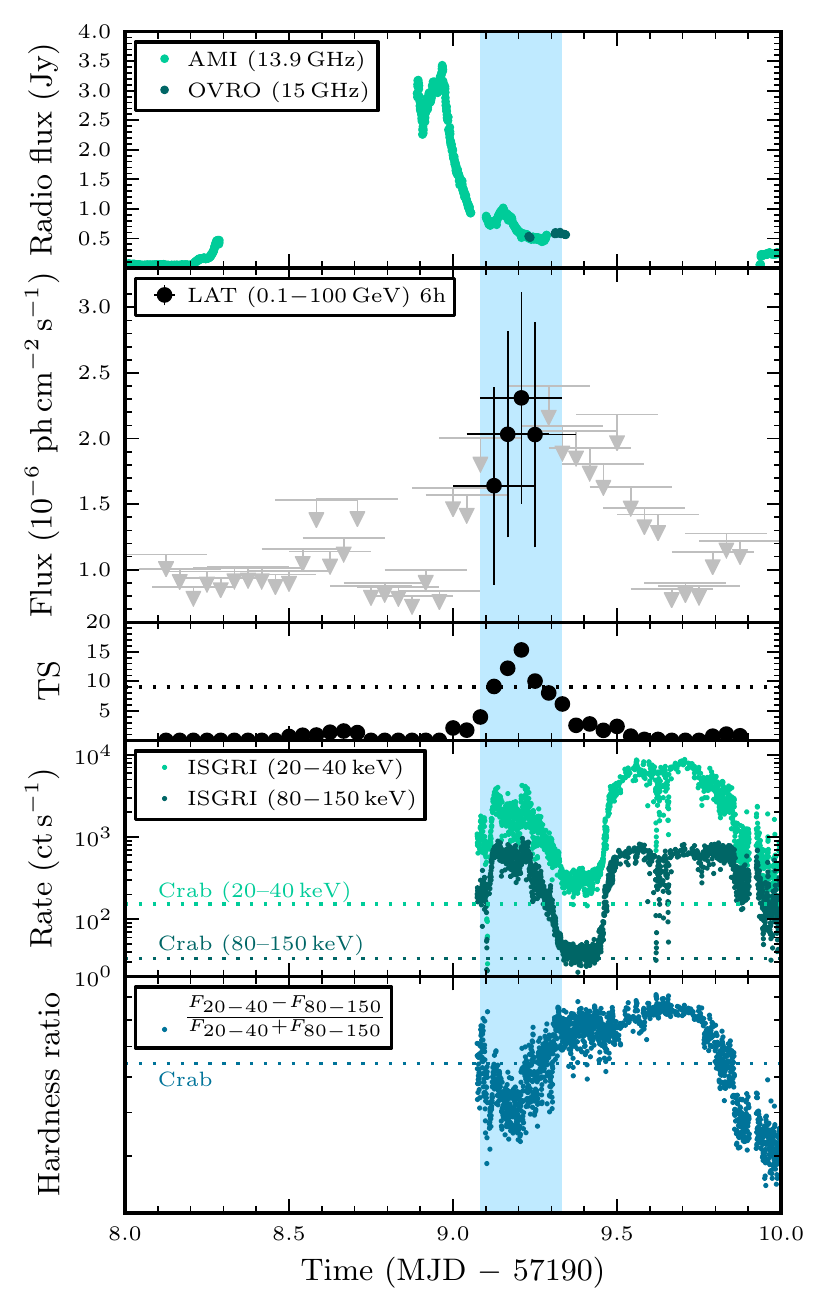}
\vspace{-20pt}
\caption{
Multi-wavelength LC of V404$\,$Cyg around the time of the $\gamma$-ray peak (vertical blue shaded area). From top to bottom panel: AMI and OVRO flux density evolutions; LAT flux and $95\%$ upper limits in 6-h bins, with 1-h shifts in exposure between each measurement; corresponding TS; INTEGRAL/ISGRI (LC and hardness ratio).
\label{fig:lc_mw}}
\end{figure}

To study the shortest time-scale variations, we have fixed every source model parameter at their fitted values from the \texttt{binned} analysis and added a point-source at the location of V404$\,$Cyg, modelled as a power-law with free normalisation and index. 
\texttt{Unbinned} analyses over $12$ or $6$-h bins were performed, with each bin shifted by $6$ or $1$-h from the previous one respectively (see Fig.~\ref{fig:lc_12h} and \ref{fig:lc_mw}).
Whenever the derived Test Statistic (TS\footnote{$\rm{TS}= 2 \ln (\mathcal{L}_{1} / \mathcal{L}_{0})$, $\mathcal{L}_{1}$ and $\mathcal{L}_{0}$ are the likelihood maxima with or without including the target source into the model.}) value was lower than $9$ \citep[corresponding to ${\lesssim} 3\sigma$ detection significance,][]{1996ApJ...461..396M}, we computed $95$ per cent upper limits on the high-energy flux using the method of \citet{1991NIMPA.300..132H} as implemented in the \texttt{pyLikelihood} module of the Science Tools. 
Otherwise, integrated $\gamma$-ray fluxes along with $1\sigma$ statistical error bars are provided. 
The LAT exposure (averaged on $12$\,h) ranges between $1$ and $4\times 10^7$\,cm$^2$\,s during the considered period, anti-correlated with the flux upper limits in Fig.~\ref{fig:lc_12h}, and is ${\sim}2.5\times 10^7$\,cm$^2$\,s during the $\gamma$-ray excess on MJD 57199.2.

Our variability analysis (performed between MJD 57140 and 57225) encompassing the June outburst period of V404$\,$Cyg reveals the presence of a weak $\gamma$-ray excess at a location close to the LMXB around MJD 57199 (26 June 2015). 
The date of the $\gamma$-ray excess maximum statistical significance is estimated at MJD $57199.2\pm 0.1$ (based on our 6-hour bins, see Fig.~\ref{fig:lc_mw}), although we caution that the measurements are not independent since the bins overlap in exposure. 
At the position of V404\,Cyg, we measure a peak photon flux $F_\gamma=(2.3 \pm 0.8) \times 10^{-6}\,$ph$\,$cm$^{-2}\,$s$^{-1}$, using the $6$-hour bins (quoted errors are statistical and dominate the systematic uncertainties), for a corresponding TS value of $15.3$ 
(the maximum on $12$-hour bins is $F_\gamma=(1.4 \pm 0.5) \times 10^{-6}\,$ph$\,$cm$^{-2}\,$s$^{-1}$ for ${\rm TS} = 15.2$, see Fig.~\ref{fig:tsmap}).
The spectrum is very soft and scales as $F_{\gamma} \propto E^{-3.5\pm0.8}$. 

As a verification procedure, we have computed the $12$-h bin lightcurve (LC) of the closest \textit{Fermi} source: 3FGL J2025.2$+$3340, with the normalisation of its spectral component let free to vary and excluding V404$\,$Cyg from the model. This blazar underwent one flaring episode in July 2009 which lasted tens of days \citep[\textit{Fermi} All-sky Variability Analysis,][]{2013ApJ...771...57A} and is labelled as variable in the 3FGL catalogue \citep[with a variability index of $278.7$,][]{2012ApJ...746..159K, 2015ApJ...810...14A}. Its LC shows no significant detection over the $30$~days encompassing the outburst period of the LMXB. 
The highest value (${\rm TS} {\sim} 8.1$) occurs on MJD~57199.25, which is consistent with the observed $\gamma$-ray excess since some of the photons are attributed to the blazar's emission if V404\,Cyg is not taken into account. Furthermore, we note that no multi-wavelength activity has been reported from this source despite intense monitoring of this field at all wavelengths in June 2015, including its use as a phase calibrator for V404\,Cyg radio observations.

	\subsection{Test Statistic maps and localisation}\label{sec:tsmap}

\begin{figure}
\includegraphics[scale=1]{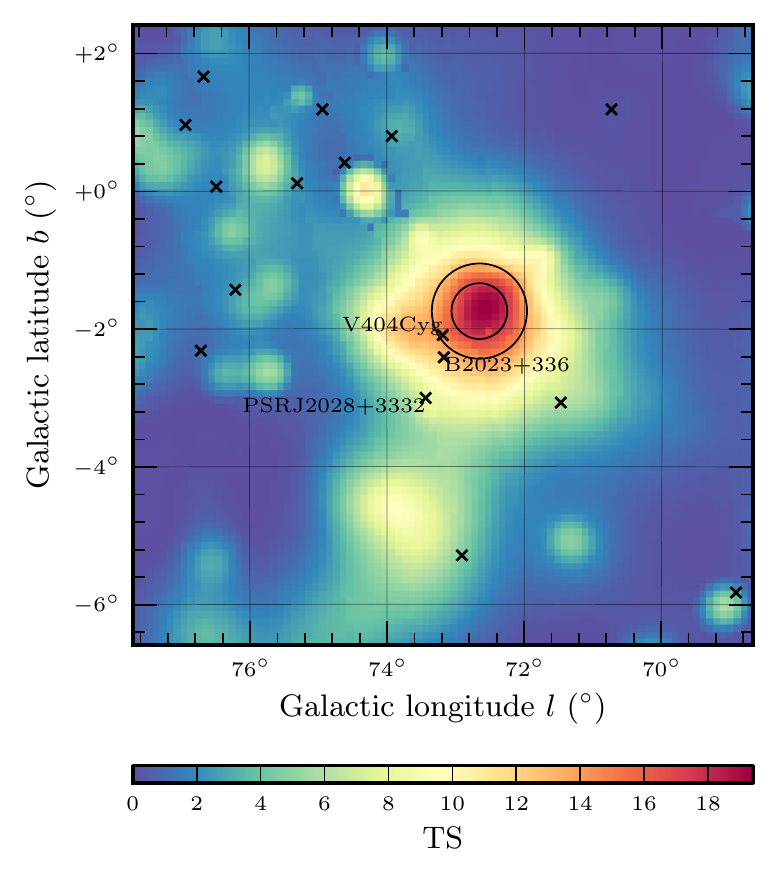}
\vspace{-10pt}
\caption{$12$-h residual TS map on MJD~$57199.25$ ($0.1$--$2\,$GeV, $0\fdg1$\,pixel$^{-1}$). 
3FGL sources are marked as black crosses. $68\%$ and $95\%$ confidence regions on the best fit position are represented. 
\label{fig:tsmap}}
\end{figure}

We constructed TS maps during the outburst of V404\,Cyg to test whether the detection could be due to a nearby source. We show in Fig.~\ref{fig:tsmap} the TS map built from a $12$-h integration around the date of the highest flux in the $0.1$--$2\,$GeV energy range and excluding V404\,Cyg from the source model. The residual TS map shows a transient excess close to the location of V404\,Cyg with a peak TS value of about $20$ (${\sim} 4.5 \sigma$). 
The TS value drops to ${\sim}6$ if photons below $200\,$MeV are not considered, in agreement with the computed soft spectrum.

We used the tool \texttt{gtfindsrc} to localize the point source at the origin of the $\gamma$-ray excess emission.
The best-fit location is at ${\rm RA(J2000)}=305\fdg31$, ${\rm Dec.}=33\fdg63$ with a TS of $17.6/19.5$ on $6/12\,$h and a similar flux than reported in \S\ref{sec:analysis_lc}. 
The $68$ and $95$ per cent confidence radii are ${\rm r68}=0\fdg43$ and ${\rm r95} =0\fdg69$ (including corrections, 3FGL), represented as concentric circles in Fig.~\ref{fig:tsmap}.  
The excess lies $0\fdg63$ from V404\,Cyg's position, $0\fdg84$ away from the blazar B2023$+$336 and $1\fdg48$ away from the next closest source, the pulsar PSR~J2028$+$3332 \citep[3FGL~J2028.3$+$3332,][]{2012ApJ...744..105P}. 
The latter is not known to vary (variability index of $51.2$, 3FGL). The position of the excess $\gamma$-ray emission thus excludes the two closest known sources. However, V404\,Cyg is within the $95$ per cent confidence region of the excess. The localization cannot be better constrained due to the softness of the photons, implying a larger PSF, and the short integration time, limiting the number of photons.

	\subsection{Hard X-ray and radio observations}
In Fig.~\ref{fig:lc_mw} we plot the V404\,Cyg LCs obtained from observations with the IBIS/ISGRI imager on board the INTEGRAL satellite \citep[data reduction similar to][]{2015A&A...581L...9R}. 
$100$-s binned LCs and hardness ratios are presented in the bands $20$--$40$ and $80$--$150$\,keV along with mean Crab levels over the year 2015 (respectively $152.9$ and $33.2\,$ct$\,$s$^{-1}$).

The $15\,$GHz observations were obtained using the Owens Valley Radio Observatory (OVRO) $40$-m telescope. Calibration is achieved using a temperature-stable diode noise source to remove receiver gain drifts and the flux density scale is derived from observations of 3C\,286 \citep{1977A&A....61...99B}. 
Details of the reduction and calibration procedure are found in \citet{2011ApJS..194...29R}. 
V404\,Cyg was observed with the Arcminute Microkelvin Imager (AMI-LA, $13$--$17\,$GHz), operating as part of the University of Oxford's 4 PI SKY transients programme (Fender et al., in prep.). 
J2025$+$3343 was used as phase calibrator and the absolute flux calibration was done using 3C\,286 \citep{1977A&A....61...99B}. 
The calibration and RFI excision were done using \texttt{AMI-REDUCE} \citep{2009MNRAS.400..984D}. 
LCs centered at $13.9\,$GHz were extracted from the calibrated data via vector-averaging of the UV data.

\section{Discussion} \label{sec:discussion}

We have found transient excess $\gamma$-ray emission at a position consistent with that of V404\,Cyg, which we could not associate with a previously-known {\it Fermi}/LAT source. Formally, the TS of this excess is not sufficiently high to claim a detection, once the number of trials associated with constructing the LC is taken into account. 
The probability of measuring a TS (distributed as a $\chi^2_2$) value $\geq 15.3$ over the 10 days of the outburst, given 320, $6$\,h-long, independent trials is ${\sim}2\%$.
The integrated PSF of the LAT for the soft excess seen toward V404\,Cyg, in the $0.1$--$2$\,GeV band, has a $1\fdg5$ HWHM. By comparing to $12$-hour-long all-sky maps averaged before and after the time interval of the X-ray outburst, we estimate the chance probability of having a transient or fluctuation as bright as the excess at ${\sim}2\%$.
So, the ${\sim}4 \times 10^{-4}$ probability of having a random, point-source, $\gamma$-ray excess at the time of the X-ray outburst, toward V404\,Cyg supports the detection of an associated $\gamma$-ray flare.
Its temporal coincidence with the brightest daily averaged \textit{Swift}/BAT flare, associated with a marked  change in the multi-wavelength properties of the source, provides a compelling reason to posit a detection of transient $\gamma$-ray emission from V404\,Cyg.

	\subsection{Multi-wavelength behaviour} \label{sec:mw_picture}

V404\,Cyg showed strong flaring activity from radio to hard X-rays over a period of about 10 days following the initial detection of the outburst on MJD 57188. 
The brightest flares occurred at the end of this period, close in time to the detection of the $\gamma$-ray excess we report. 
As shown by the AMI $13.9\,$GHz LC (Fig.~\ref{fig:lc_mw}), a giant radio flare with a peak above $3.4\,$Jy occurred ${\sim} 6\,$h before the $\gamma$-ray flare peak. 
Our $15\,$GHz OVRO data (Fig.~\ref{fig:lc_mw}), as well as VLA and VLBA  (J.~Miller-Jones, priv. com.) and RATAN \citep{2015ATel.7716....1T} observations conducted around MJD 57198 and 57200 do not indicate the presence of other major radio flares. 
At hard X-rays, the brightest flare reaches about $57\,$Crab at $20$--$40\,$keV and occurs ${\sim}14\,$h after the radio flare (i.e.\ starting on MJD 57199.5). 
This is preceded by another hard X-ray flare around MJD 57199.2 that is coincident with the \textit{Fermi}/LAT $\gamma$-ray excess. 
Although its $20$--$40\,$keV intensity is lower than the MJD 57199.5 flare, its $80$--$150\,$keV flux is higher (and corresponds to the maximum at ${\sim}29\,$Crab over the whole outburst period, see Fig.~\ref{fig:lc_mw}). 
The activity on MJD 57198--57200 in radio and hard X-rays is also accompanied by the third (and last) detection of $e^-e^+$ pair annihilation \citep{2016Natur.531..341S}. 
The annihilation flux LC follows the $100$--$200\,$keV flux over these two days, including the dip at MJD 57199.4. 
The X-ray spectral fits suggest annihilation occurs in two zones with, respectively, $kT {\sim} 2\,\rm keV$ and ${\sim} 500\,\rm keV$. 
\citet{2016Natur.531..341S} performed a LAT study during the largest positron flare (MJD 57199.616--57200.261) and derived a $8 \times 10^{-7}\,$ph$\,$cm$^{-2}\,$s$^{-1}$ upper limit that is consistent with our upper limit of $8.7 \times 10^{-7}\,$ph$\,$cm$^{-2}\,$s$^{-1}$ on the same time interval, above $100$\,MeV.
The activity and emission levels decrease markedly at all wavelengths after MJD 57200. 
The $\gamma$-ray excess thus appears related to the last spur of activity of the source, preceding the brightest enhancement of $511$\,keV emission, before it started to fade.

	\subsection{Gamma-ray emission from the jet?}
The association of the $\gamma$-ray excess with major radio activity and marked transition to a softer X-ray state (Fig.~\ref{fig:lc_mw}) are reminiscent of Cyg X$-$3, where $\gamma$-ray emission is detected when the radio flux is above ${\sim} 0.2\,$Jy and when X-ray emission is soft -- but not too much so \citep{2012MNRAS.421.2947C}. 
The $\gamma$-ray luminosity from V404\,Cyg is ${\sim} 2\times 10^{35} (d/{\rm 2.4\,kpc})^2\,\rm erg\,s^{-1}$, a factor ${\sim} 5$ fainter than the typical ${>}100$\,MeV luminosity from Cyg X$-$3. 
This is a minor fraction of the overall luminosity output, which reached  ${\sim} 2\times 10^{38}\,\rm erg\,s^{-1}$ \citep{2015A&A...581L...9R}. 
Gamma-ray emission may thus be much fainter in V404\,Cyg than in Cyg X$-$3, either intrinsically or because of the boosting depending on the Lorentz factor and orientation of the emission in the relativistic jet \citep{2005Ap&SS.297..109G}.  
Radio VLBI imaging of the jet should be able to narrow down such a possibility.
In both cases, it is likely that we have detected only the brightest flaring episode during the outburst.

V404\,Cyg appeared to stay in a hard or intermediate state during this outburst \citep{2015A&A...581L...9R, 2016arXiv160100911J, 2016arXiv160103234R, 2016Natur.529...54K}.
The tentative high-energy $\gamma$-ray detections of the microquasar Cyg X-1 are also associated with this spectral state \citep[and/or at the transition to the soft state,][]{2013ApJ...766...83S}. 
\citet{2013ApJ...775...98B} found 21 daily measurements for which $\rm{TS}>9$ at the position of the source, mostly when it was in a hard or intermediate state. 
\citet{2013MNRAS.434.2380M} obtained a total $\mathrm{TS} {\sim} 15$ by integrating all the hard state data and no detection in the soft state. 
The $\gamma$-ray emission from Cyg X$-$1 is not a simple extrapolation of the ${>}100\,$keV power-law emission detected in X-rays, associated with non-thermal emission from the corona and/or the jet \citep{2015ApJ...807...17R}. 
Models suggest that the high-energy $\gamma$-ray emission is located elsewhere, most likely associated with non-thermal emission from the jet \citep{2013MNRAS.434.2380M}. 
The association in V404\,Cyg of high-energy $\gamma$-ray emission contemporaneous with annihilating $e^+e^-$ pairs further supports this picture. 
Radiative models of coronal plasmas show significant $511\,$keV line emission when the pair energy input is primarily non-thermal and when the corona is very dense i.e. a large opacity to pair production on ${>}100\rm\,MeV$ photons \citep[e.g.][]{1987MNRAS.227..403S}. 
The high-energy $\gamma$-ray emission must thus originate away from the corona.

An intriguing possibility is that the ejection during the radio flare is associated with the disruption of a jet extracting energy from the black hole through the Blandford-Znajek process \citep{1977MNRAS.179..433B}. 
Numerical simulations indicate this process to be efficient when magnetic flux dragged in by a thick accretion flow piles up to create a `magnetically-arrested disk' (MAD) close to the black hole \citep{2011MNRAS.418L..79T, 2012MNRAS.423.3083M}. 
\citet{2016ApJ...819...95O} show that such a configuration leads to high-energy $\gamma$-ray emission in the hard state, due to Compton upscattering of jet or disk synchrotron photons. 
In contrast, the $\gamma$-ray emission is very weak when the built-up magnetic flux is insufficient to suppress standard accretion. 
In this scenario, a transient ejection occurs when the MAD configuration is disrupted (reconnects) due to incoming magnetic flux of opposite polarity \citep{2014MNRAS.440.2185D}. 
\citet{2016ApJ...819...95O} indicate that the re-formed jet is brighter in $\gamma$-rays than prior to the major ejection. 
Hence, the brief $\gamma$-ray emission from V404 Cyg following a major radio flare might be associated with the temporary re-formation of a powerful Blandford-Znajek jet.

The present observations support the presence of $\gamma$-ray emission associated with ejections in microquasars, and for the first time from an accreting black hole with a low-mass donor companion. 
This emission is weak and variable, with a low statistical significance. 
Some perseverance will be required to clarify the multi-wavelength context in which it appears in this and other microquasars, thus fulfilling its potential as a diagnostic of the accretion-ejection process.

\section*{Acknowledgements}
{\small
We thank J. Miller-Jones, J. Ballet and J. Perkins for useful discussions and T. Siegert for providing the annihilation LC.
AL, SC and JR acknowledge funding support from the French Research National Agency: CHAOS project ANR-12-BS05-0009 and the UnivEarthS Labex program of Sorbonne Paris Cit\'e (ANR-10-LABX-0023 and ANR-11-IDEX-0005-02).
KM acknowledges funding support from the Hintze Foundation.
The \textit{Fermi}-LAT Collaboration acknowledges support for LAT development, operation and data analysis from NASA and DOE (United States), CEA/Irfu and IN2P3/CNRS (France), ASI and INFN (Italy), MEXT, KEK, and JAXA (Japan), and the K.A.~Wallenberg Foundation, the Swedish Research Council and the National Space Board (Sweden). Science analysis support in the operations phase from INAF (Italy) and CNES (France) is also gratefully acknowledged.
The OVRO 40-m monitoring program is supported in part by NASA grants NNX08AW31G and NNX11A043G, and NSF grants AST-0808050 and AST-1109911.
We acknowledge with thanks the variable star observations from the AAVSO.
}



\bibliographystyle{mnras}
\bibliography{loh_v404cyg} 



\label{lastpage}
\end{document}